\documentclass[aps,prd,twocolumn,groupedaddress,showpacs,superscriptaddress,floatfix]{revtex4-1}

\usepackage{amsmath}\allowdisplaybreaks[1]
\usepackage{graphicx}
\usepackage{amssymb}
\usepackage{bm}
\usepackage{array}
\usepackage[colorlinks,linkcolor=blue,urlcolor=blue,anchorcolor=blue,citecolor=blue]{hyperref}
\usepackage{slashed}
\usepackage{braket}
\usepackage{xcolor}

\begin{document}

\title{Baryon properties from light-front holographic QCD}
\preprint{Published in Phys.Rev. D92 (2015) 096003.}

\newcommand*{\PKU}{School of Physics and State Key Laboratory of Nuclear Physics and Technology, Peking University, Beijing 100871,China}\affiliation{\PKU}
\newcommand*{\CICQM}{Collaborative Innovation Center of Quantum Matter, Beijing, China}\affiliation{\CICQM}
\newcommand*{\CHEP}{Center for High Energy Physics, Peking University, Beijing 100871, China}\affiliation{\CHEP}

\author{Tianbo Liu}\email{liutb@pku.edu.cn}\affiliation{\PKU}
\author{Bo-Qiang Ma}\email{mabq@pku.edu.cn}\affiliation{\PKU}\affiliation{\CICQM}\affiliation{\CHEP}

%\date{\today}

\begin{abstract}

We investigate the properties of octet and decuplet baryons in a light-front holographic model. By taking into account the effect of nonvanishing quark mass, we obtain the modified light-front wave functions which are applicable at both low and high energy scales. We calculate the spectra, form factors, magnetic moments and electromagnetic radii of octet and decuplet baryons with the results all matching the experiments well. The axial charge, which describes the contribution of quark helicity to the proton spin in the quark-parton model at the high energy scale, is also consistent with the experimental value. Therefore, the light-front holographic method is successful in studying hadronic physics at all energy scales, and the nonzero quark mass is essential to understand the spin structures together with other low energy properties.

\end{abstract}


\maketitle

Quantum chromodynamics (QCD) provides a fundamental description of
strong interaction in terms of quark and gluon degrees of freedom,
and has been proven successful in the high energy region where the perturbative
effect dominates. It is still challenging to directly calculate the hadron properties at the low energy
scale where the strong and nonlinear coupling between quarks and gluons plays the essential role.
The light-front holography provides a new method to handle the strong interaction from basic considerations and has led to many remarkable results~\cite{Brodsky:2014yha}. The light-front holography is
based on the correspondence~\cite{Maldacena:1997re,Gubser:1998bc,Witten:1998qj}
between string states defined on the five-dimensional anti-de Sitter
(AdS) space-time and conformal field theories (CFT) in physical
space-time.
Although the conformal symmetry of the classical QCD Lagrangian with massless quarks is broken by quantum effects, it is nearly conformal at high energy or short distance because of its asymptotic freedom~\cite{Gross:1973ju}. There is also evidence from lattice QCD~\cite{Furui:2006py}, the Dyson-Schwinger equation~\cite{von Smekal:1997is}, empirical effective charges~\cite{Deur:2005cf} and theoretical arguments~\cite{Brodsky:2008be} that the strong coupling constant has an infrared fixed point. Therefore, the AdS/CFT correspondence can be used to obtain a first approximation to QCD. This kind of methods have been applied to the hadronic spectrum~\cite{Karch:2006pv}, hard scattering~\cite{Polchinski:2001tt} and strongly coupled quark-gluon plasma~\cite{CasalderreySolana:2011us}.

During the last decade, a connection is established between the conformal quantum mechanics and the light-front dynamics~\cite{Brodsky:2003px,Brodsky:2013ar}, which provides a natural framework to reconcile the quark-parton model with QCD~\cite{Brodsky:1997de,Dirac:1949cp}. The simple vacuum in the light-front quantization allows an unambiguous definition of the constituents of hadrons. All hadronic properties and partonic structures are encoded in the frame independent light-front wave functions (LFWFs). Therefore, solving the LFWFs becomes a central as well as challenging issue in hadronic physics. Recently, an exact correspondence is found between the fifth-dimensional coordinate $z$ in AdS space and the weighted impact separation variable $\zeta$ in physical space-time~\cite{Brodsky:2007hb,Brodsky:2008pf}. This endows a clear physical meaning to the holographic variable. Some LFWFs are derived from some holographic models as a first approximation~\cite{Brodsky:2006uqa,deTeramond:2008ht}.

In this paper, we study the baryon properties with the light-front holographic approach. Including the effects from nonzero quark mass, which explicitly breaks the conformal symmetry but plays an important role in understanding spin-related issues, we obtain the LFWFs with different orbital angular momenta and find that the spectra, magnetic moments, form factors and electromagnetic radii of octet and decuplet baryons are all well described. It is remarkable that the axial charge which describes the fraction of the quark helicity contribution to the proton spin is also consistent with the experimental value, when the quark mass effect is taken into account. Therefore, the new LFWFs we obtained are applicable to study the baryon properties at both low energy and high energy scales.

Hadrons, as bound states of the strong interaction, are the eigenstates of the QCD light-front Hamiltonian $H_{\textrm{LF}}=2P^+P^--\bm{P}_\perp^2$ with mass squares as the eigenvalues. Quantized at fixed light-front time, it can be expanded on the Fock state basis as
\begin{equation}
|H\rangle=\sum_n \int[dx][d^2\bm{k}_\perp]\psi_{n/H}(x_i,\bm{k}_{i\perp})|n:x_i,\bm{k}_{i\perp},\lambda_i\rangle,
\end{equation}
where the integral measures are defined as
\begin{align}
\int[dx]&=\prod_{i=1}^n\int dx_i\delta(1-\sum_{j=1}^n x_j),\\
\int[d^2\bm{k}_\perp]&=\prod_{i=1}^n\int \frac{d^2\bm{k}_{i\perp}}{2(2\pi)^3}16\pi^3\delta^{(2)}(\sum_{j=1}^n\bm{k}_{j\perp}),
\end{align}
and $\lambda_i$ represents the helicity and other internal degrees of freedom. Since there is an explicit separation of kinematical and dynamical terms in QCD light-front Hamiltonian, we can express the mass square of a hadron in terms of the LFWFs as
\begin{align}
M_H^2&=\int[dx][d^2\bm{k}_\perp]\sum_{i=1}^n\frac{\bm{k}_{i\perp}^2+m_i^2}{x_i}|\psi(x_i,\bm{k}_{i\perp})|^2\nonumber\\
&\quad +\int[dx][d^2\bm{k}_\perp]\psi^*(x_i,\bm{k}_{i\perp})U\psi(x_i,\bm{k}_{i\perp}),\label{mass2}
\end{align}
where $U$ is an effective potential. In this semiclassical approximation, some nondiagonal effects are neglected.

Using the Fourier transformation, one may obtain the expression in the coordinate space. For a two-body system, which means the hadron is regarded as an active quark and a spectator cluster, the eigenequation with massless constituents is expressed as~\cite{deTeramond:2008ht}
\begin{equation}\label{lfequation}
\bigg(-\frac{d^2}{d\zeta^2}-\frac{1-4L^2}{4\zeta^2}+\tilde{U}\bigg)\phi(\zeta)=M_H^2\phi(\zeta),
\end{equation}
where the light-front variable $\zeta=\sqrt{x(1-x)}|\bm{b}_\perp|$ measures the separation between the quark and the spectator. It corresponds to the holographic variable $z$ in AdS space. The transverse mode $\phi(\zeta)$ of the LFWF is defined as~\cite{Brodsky:2007hb}
\begin{equation}
\psi(x,\bm{b}_\perp)=e^{iL\varphi}X(x)\frac{\phi(\zeta)}{\sqrt{2\pi\zeta}}.
\end{equation}

The effective potential $\tilde{U}$ is usually derived from the deformation of the AdS space, i.e., the soft-wall model~\cite{Karch:2006pv}, by introducing a dilaton $\varphi(z)=\lambda z^2$. For fermions, however, the dilaton can always be absorbed through the rescaling of the fields~\cite{Kirsch:2006he}. Therefore, an effective interaction $\rho(z)$ is introduced to the effective action in AdS space~\cite{deTeramond:2013it},
\begin{equation}
S_{\textrm{eff}}=\int d^4xdz\sqrt{g}e^{\varphi(z)}\bigg[\bar{\Psi}\big(i\Gamma^Ae_A^MD_M-\mu-\rho(z)\big)\Psi+h.c.\bigg],
\end{equation}
where $\Psi$ is the bulk field, $\Gamma^A$ is the tangent-space Dirac matrices, and $e^M_A$ is the inverse vielbein.

The bayron light-front wave function satisfies the coupled linear differential equations~\cite{Brodsky:2008pg},
\begin{align}
\bigg(-\frac{d}{d\zeta}-\frac{\nu+\frac{1}{2}}{\zeta}-V(\zeta)\bigg)\phi_-&=M\phi_+,\label{ode1}\\
\bigg(\frac{d}{d\zeta}-\frac{\nu+\frac{1}{2}}{\zeta}-V(\zeta)\bigg)\phi_+&=M\phi_-,\label{ode2}
\end{align}
where the subscripts $\pm$ represent the chiral components. The confinement potential $V(\zeta)$ is determined by the effective interaction $\rho(z)$ as
\begin{equation}
V(\zeta)=\frac{R}{\zeta}\rho(\zeta).
\end{equation}
In this study, we choose a linear potential $V(\zeta)=\lambda\zeta$ which can reproduce the Regge behavior for bayrons~\cite{Brodsky:2008pg,Abidin:2009hr}. This form can also be uniquely determined in the framework of superconformal algebra~\cite{deTeramond:2014asa}. From the coupled differential equations \eqref{ode1} and \eqref{ode2}, one can obtain the equivalent second-order equations as
\begin{align}
\bigg(-\frac{d^2}{d\zeta^2}-\frac{1-4\nu^2}{4\zeta^2}+\tilde{U}_+(\zeta)\bigg)\phi_+&=M^2\phi_+,\\
\bigg(-\frac{d^2}{d\zeta^2}-\frac{1-4(\nu+1)^2}{4\zeta^2}+\tilde{U}_-(\zeta)\bigg)\phi_-&=M^2\phi_-,
\end{align}
where the effective potential is expressed as
\begin{equation}
\tilde{U}_\pm(\zeta)=V^2(\zeta)\pm V'(\zeta)+\frac{1+2\nu}{\zeta}V({\zeta}).
\end{equation}
Comparing with \eqref{lfequation}, one may relate $\nu$ to the orbital angular momentum $L$. To have the separation between the kinematical and dynamical effects, we assign $L=\nu$ for the right-hand component and $L=\nu+1$ for the left-hand component.

We take the lowest energy solutions with $\nu=0$ for the ground state baryons,
\begin{align}
\phi_+(\zeta)&\sim\sqrt{2\zeta}e^{-\frac{\lambda\zeta^2}{2}}, & \phi_-(\zeta)&\sim 2\zeta^{\frac{3}{2}}e^{-\frac{\lambda\zeta^2}{2}}.
\end{align}
To include the correction from nonzero masses of the constituents, we adopt the replacement~\cite{Brodsky:2014yha},
\begin{equation}
\frac{\bm{k}_\perp^2}{x(1-x)}\rightarrow \frac{\bm{k}_\perp^2+m_1^2}{x}+\frac{\bm{k}_\perp^2+m_2^2}{1-x},
\end{equation}
to the LFWFs in the momentum space, where $m_1$ and $m_2$ are the effective masses of the quark and the spectator cluster respectively. Then the LFWFs have the form
\begin{align}
\psi_+(x,\bm{k}_\perp)&\sim \frac{4\pi}{\sqrt{\lambda x (1-x)}} e^{-\frac{1}{2\lambda}(\frac{\bm{k}_\perp^2+m_1^2}{x}+\frac{\bm{k}_\perp^2+m_2^2}{1-x})},\label{l0}\\
\psi_-(x,\bm{k}_\perp)&\sim \frac{4\pi|\bm{k}_\perp|}{\lambda x (1-x)} e^{-\frac{1}{2\lambda}(\frac{\bm{k}_\perp^2+m_1^2}{x}+\frac{\bm{k}_\perp^2+m_2^2}{1-x})}.\label{l1}
\end{align}
They actually describe $L=0$ and $L=1$ states respectively. In other words, the ground state bayrons contain the nonzero orbital angular momentum state in the light-front dynamics. This is a relativistic effect that can be physically understood from the Wigner rotation effect~\cite{Wigner:1939cj}, which plays an important role in explaining the proton spin puzzle~\cite{Ma:1991xq}. On the limit of zero quark mass, there is an equal probability to find the $L=0$ and the $L=1$ states in a baryon state~\cite{deTeramond:2013it}. In such a situation, the proton spin is all from the orbital angular momentum.

However, the nonvanishing quark mass changes the weights of these two states. Comparing the ratio with the generic ansatz,
\begin{equation}
\frac{|\psi_{L=0}|^2}{|\psi_{L=1}|^2}=\frac{(m_1+x\mathcal{M})^2}{\bm{k}_\perp^2},
\end{equation}
where $\mathcal{M}$ is a quantity with the dimension of mass, we identify the $x\mathcal{M}$ with $\sqrt{\lambda x (1-x)}$ in order to reproduce the ratio between \eqref{l0} and \eqref{l1} on the limit of massless quarks. We obtain the modified LFWFs as
\begin{align}
\psi_{L=0}(x,\bm{k}_\perp)&=N\frac{4\pi[m_1+\sqrt{\lambda x(1-x)}]}{\lambda x (1-x)}\nonumber\\
&\quad\times e^{-\frac{1}{2\lambda}(\frac{\bm{k}_\perp^2+m_1^2}{x}+\frac{\bm{k}_\perp^2+m_2^2}{1-x})},\\
\psi_{L=1}(x,\bm{k}_\perp)&=N\frac{-4\pi(k^1+ik^2)}{\lambda x(1-x)}\nonumber\\
&\quad\times e^{-\frac{1}{2\lambda}(\frac{\bm{k}_\perp^2+m_1^2}{x}+\frac{\bm{k}_\perp^2+m_2^2}{1-x})},
\end{align}
where $N$ is the normalization factor. Then the baryon mass can be evaluated via \eqref{mass2} with the contributions from kinematical energy, the confinement potential and constituent masses.

In the light-front formalism, the Dirac and Pauli form factors correspond to the spin-conserving and the spin-flip current matrix elements respectively~\cite{Brodsky:1980zm}. With the electromagnetic current $V(z,Q^2)$ in AdS space, the Dirac form factor is expressed as
\begin{equation}
F_1(Q^2)=\sum_L\int dz \frac{R^4}{z^4}V(z,Q^2)\Psi_L^2(z).
\end{equation}
Here we adopt the dressed current derived from the soft-wall model~\cite{Grigoryan:2007my,Brodsky:2007hb},
\begin{equation}
V(z,Q^2)=\Gamma\left(1+\frac{Q^2}{4\lambda}\right)U\left(\frac{Q^2}{4\lambda},0,\lambda z^2\right),
\end{equation}
where $U(\alpha,\gamma,z)$ is the Tricomi confluent hypergeometric function.

Since the precise mapping of the Pauli form factor has not been carried out in holographic methods, a nonminimal electromagnetic coupling with the anomalous gauge invariant term is proposed~\cite{Abidin:2009hr},
\begin{equation}
\eta\int d^4xdz\sqrt{g}\bar{\Psi}e^A_M e^B_N [\Gamma_A,\Gamma_B]F^{MN}\Psi,
\end{equation}
where $\eta$ is an effective coupling constant. Then the Pauli form factor is expressed as
\begin{equation}
F_2(Q^2)=\eta\int dz\frac{R^3}{z^3}\bar{\Psi}_{L=0}(z)V(z,Q^2)\Psi_{L=1}(z).
\end{equation}

\begin{figure}
\begin{center}
\includegraphics[width=0.23\textwidth]{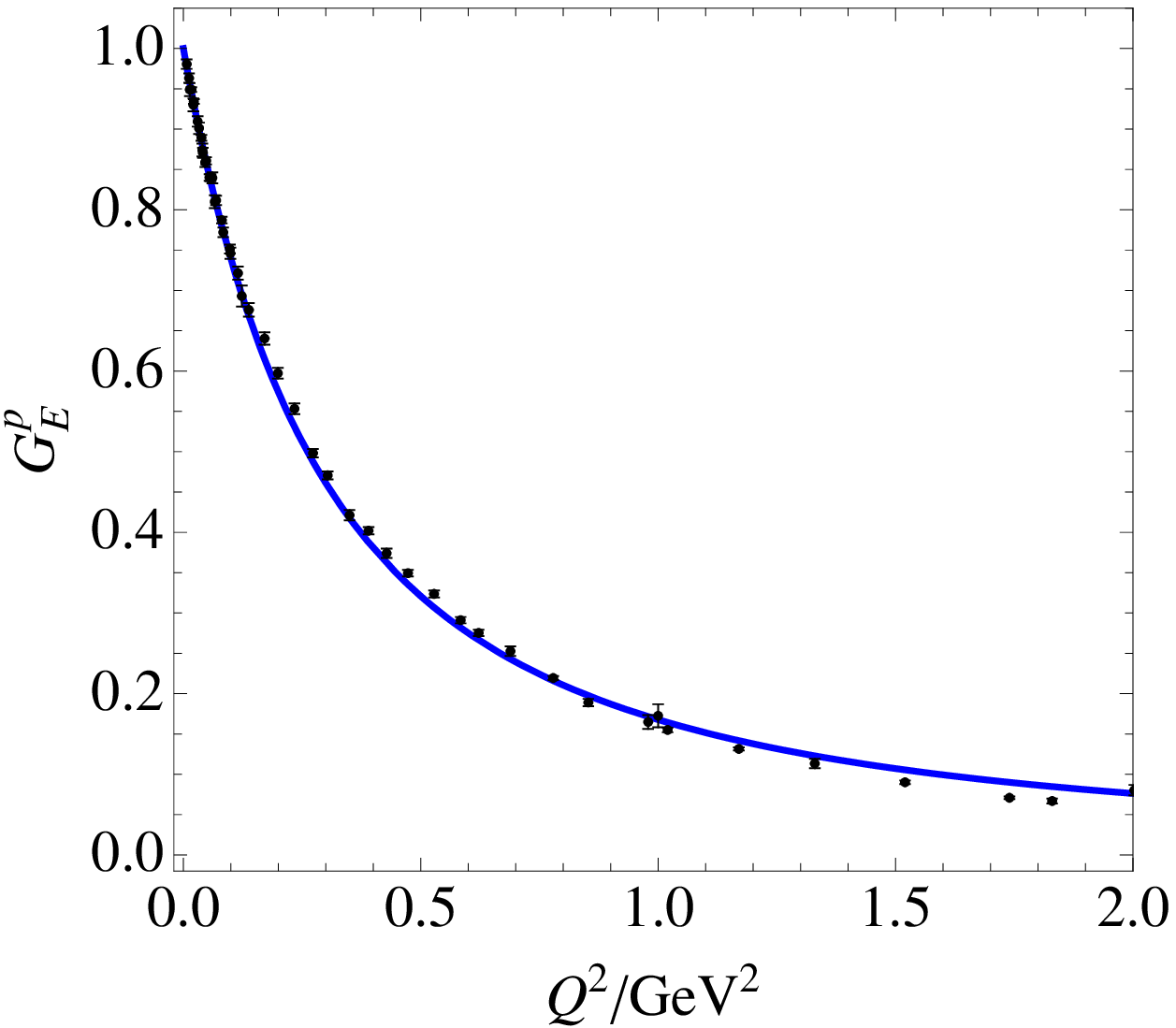}
\includegraphics[width=0.23\textwidth]{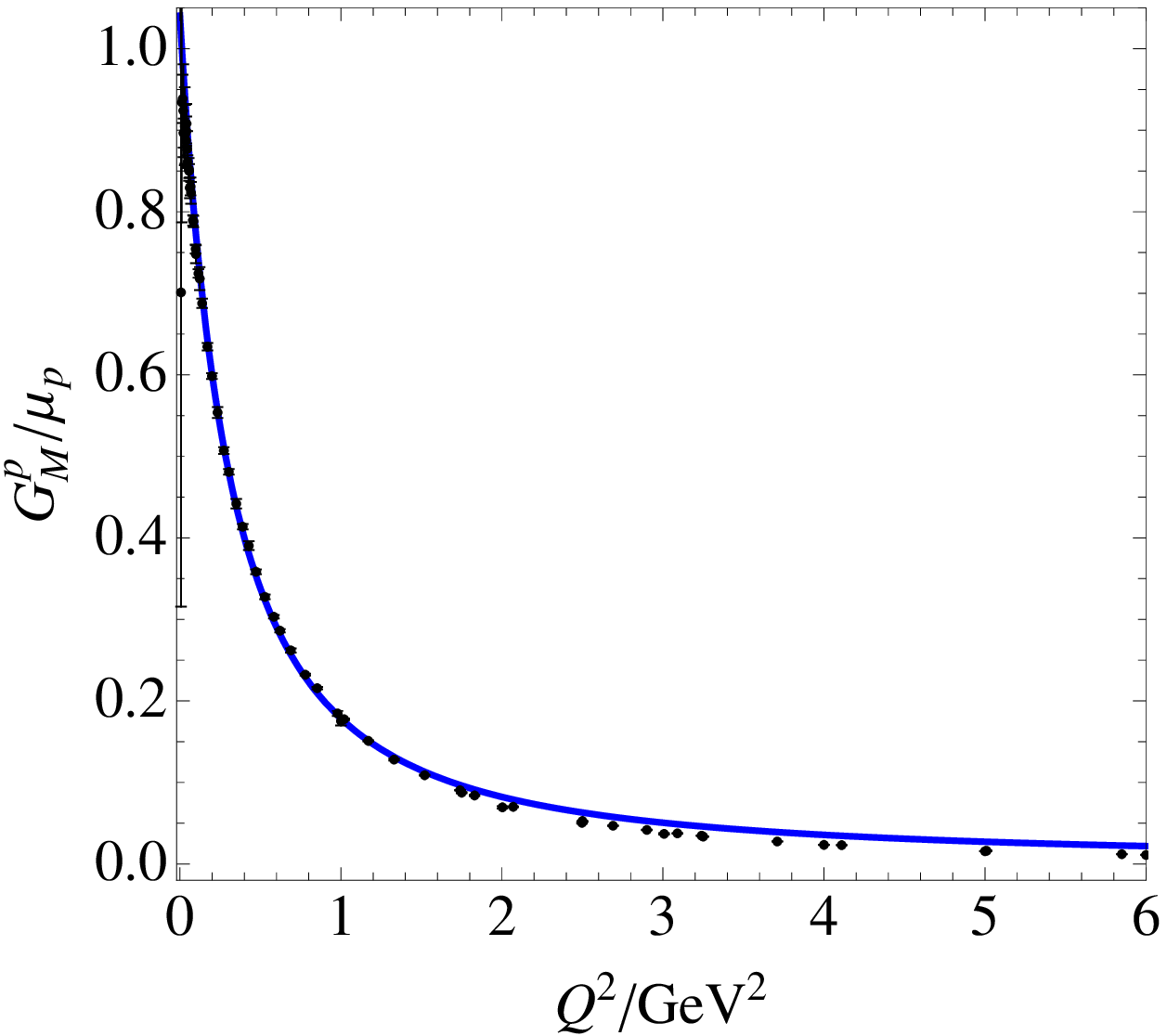}\\
\includegraphics[width=0.23\textwidth]{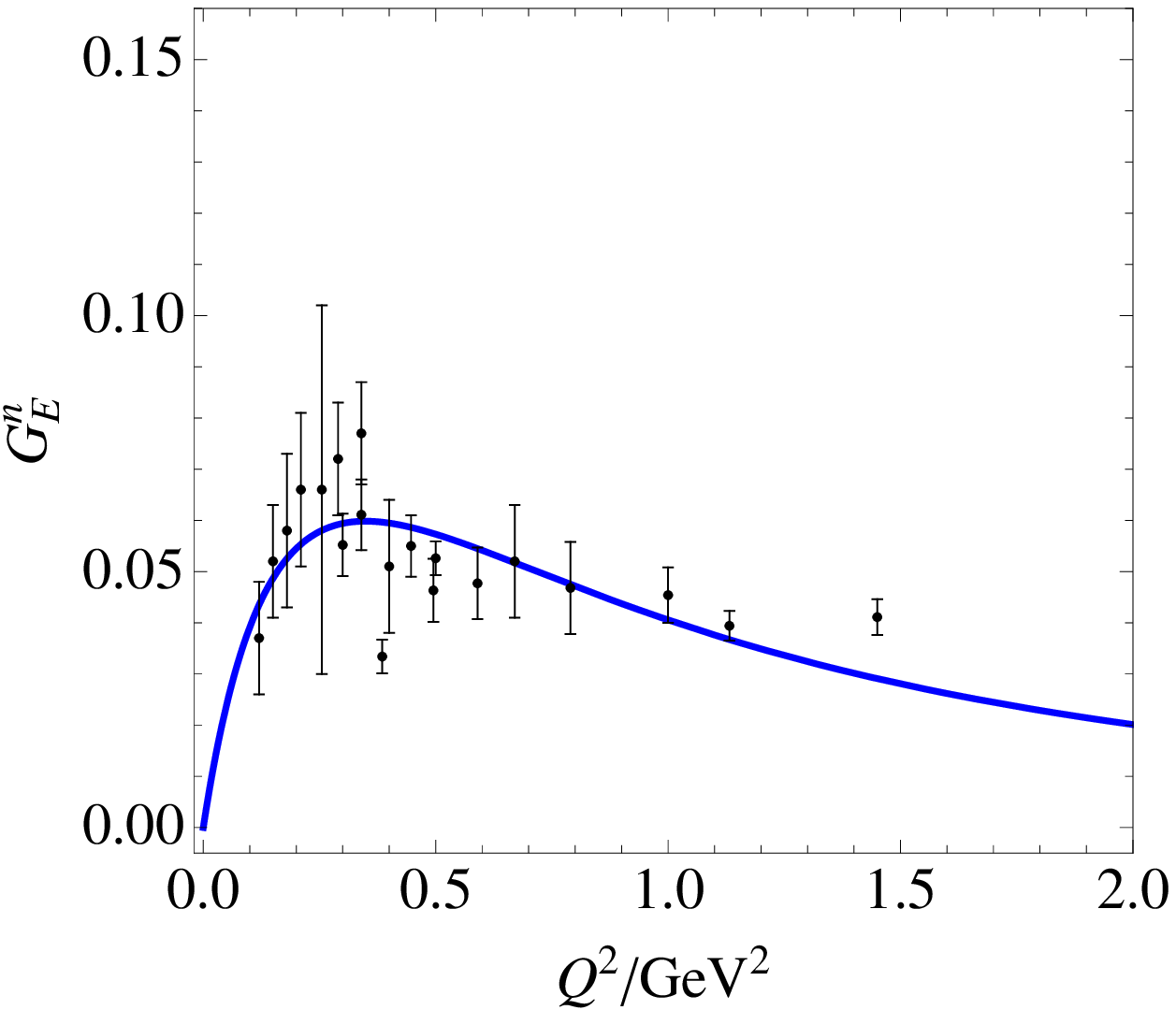}
\includegraphics[width=0.23\textwidth]{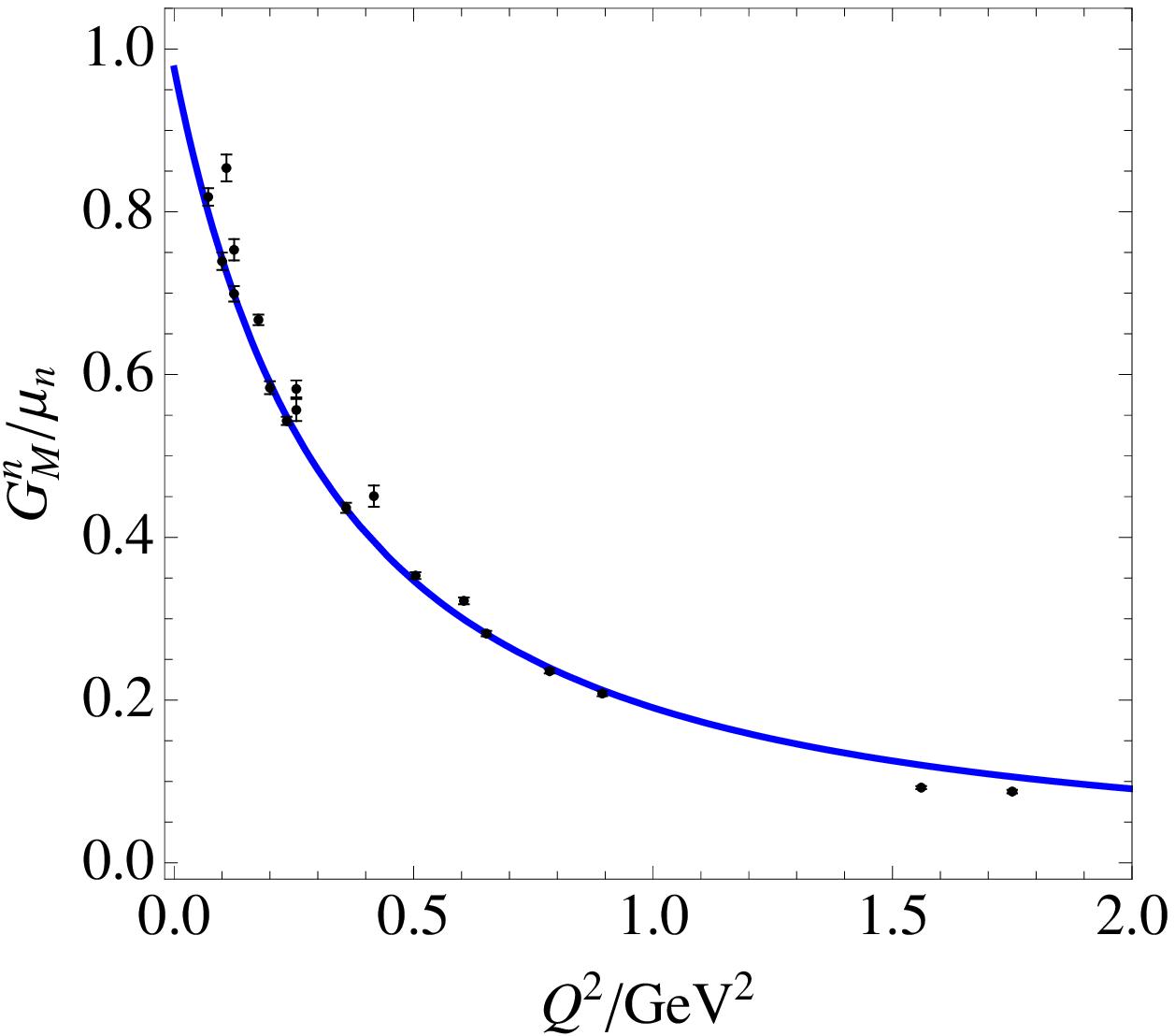}
\end{center}
\caption{The electromagnetic form factors of the proton and neutron. The curves are our model results. The data are taken from Refs.~\cite{Arrington:2007ux,Walker:1993vj,Borkowski:1974mb,Kubon:2001rj,Glazier:2004ny,Markowitz:1993hx,Bruins:1995ns,Lung:1992bu,Zhu:2001md,Warren:2003ma,Rohe:1999sh}.\label{formfactors}}
\end{figure}

The Sachs form factors are defined in terms of Dirac and Pauli form factors as~\cite{Sachs:1962zzc}
\begin{align}
G_E(Q^2)&=F_1(Q^2)-\frac{Q^2}{4M^2}F_2(Q^2),\\
G_M(Q^2)&=F_1(Q^2)+F_2(Q^2).
\end{align}
Fitted to the data, the coupling constant is chosen as $\eta=1.3$, and the parameter in the effective potential is chosen as $\lambda_{\textrm{S}}=0.105\,\textrm{GeV}^2$ for the scalar spectator cluster and $\lambda_{\textrm{V}}=0.170\,\textrm{GeV}^2$ for the vector spectator cluster. One may also use different $\eta$ for the scalar and the vector spectator cases~\cite{Abidin:2009hr}. The mass parameters are listed in Table \ref{masspara}. The results of the proton and the neutron form factors are plotted in Fig.~\ref{formfactors} compared with the experimental data taken from Refs.~\cite{Arrington:2007ux,Walker:1993vj,Borkowski:1974mb,Kubon:2001rj,Glazier:2004ny,Markowitz:1993hx,Bruins:1995ns,Lung:1992bu,Zhu:2001md,Warren:2003ma,Rohe:1999sh}.

\begin{table}
\caption{Effective masses of quarks and spectator clusters in the unit of MeV. The $q$ represents the $u$ quark and $d$ quark.\label{masspara}}
\begin{tabular*}{0.5\textwidth}{m{0.1\textwidth}c|m{0.1\textwidth}c|m{0.1\textwidth}c}
\hline\hline
~$m_q$ & $50$~~~  & ~$m_{\textrm{S}}(qq)$ & $130$~~~ & ~$m_{\textrm{V}}(qq)$ & $770$ \\
~$m_s$ & $203$~~~ & ~$m_{\textrm{S}}(qs)$ & $554$~~~ & ~$m_{\textrm{V}}(qs)$ & $1040$ \\
 & & & & ~$m_{\textrm{V}}(ss)$ & $1208$ \\
\hline\hline
\end{tabular*}
\end{table}

We calculate the spectra of octet and decuplet baryons. The results are listed in Table~\ref{spectra} compared with the experimental values in Ref.~\cite{Agashe:2014kda}. The spectator masses $m_{\textrm{S}}(qq)$, $m_{\textrm{V}}(qq)$, $m_{\textrm{S}}(qs)$, $m_{\textrm{V}}(qs)$ and $m_{\textrm{V}}(ss)$ are close to those of the mesons $\pi$, $\rho$, $K$, $K^*$ and $\phi$. The $N$-$\Delta$ mass difference is obtained from the different parameters for the scalar and the vector spectators. An alternative treatment is to adopt different assignments to $\nu$ with $L$ and $L+1/2$ for $N$ and $\Delta$ as used in Ref.~\cite{Brodsky:2014yha}.

\begin{table}
\caption{The spectra of octet and decuplet baryons.\label{spectra}}
\begin{tabular*}{0.5\textwidth}{m{0.12\textwidth}m{0.12\textwidth}rl}
\hline\hline
 Baryon & $M_{\textrm{th}}$/MeV & & $M_{\textrm{ex}}$/MeV\ \ \ \ \ \ \cite{Agashe:2014kda} \\
\hline
$N$ & 977.8 & $p$:& $938.272046 \pm 0.000021$ \\
    &       & $n$:& $939.565379 \pm 0.000021$ \\
$\Lambda$ & 1143 & $\Lambda^0$:& $1115.683 \pm 0.006$ \\
$\Sigma$ & 1146 & $\Sigma^+$:& $1189.37 \pm 0.07$ \\
         &      & $\Sigma^0$:& $1192.642 \pm 0.024$ \\
         &      & $\Sigma^-$:& $1197.449 \pm 0.030$ \\
$\Xi$    & 1349 & $\Xi^0$:& $1314.86 \pm 0.20$ \\
         &      & $\Xi^-$:& $1321.71 \pm 0.07$ \\
\hline
$\Delta$ & 1201 & $\Delta$:& $1209 \sim  1210$ \\
$\Sigma^*$ & 1370 & $\Sigma^{*+}$:& $1382.80 \pm 0.35$ \\
         &        & $\Sigma^{*0}$:& $1383.7 \pm 1.0$ \\
         &        & $\Sigma^{*-}$:& $1387.2 \pm 0.5$ \\
$\Xi^*$ & 1525 & $\Xi^{*0}$:& $1531.80 \pm 0.32$ \\
        &      & $\Xi^{*-}$:& $1535.0 \pm 0.6$\\
$\Omega$ & 1660 & $\Omega^-$:& $1672.45 \pm 0.29$ \\
\hline\hline
\end{tabular*}
\end{table}

The magnetic moment can be defined from the magnetic form factor on the limit of the zero momentum square transferred as $\mu=G_M(0)$. Our results of magnetic moments of octet baryons are listed in Table~\ref{moments} compared with the experimental values in Ref.~\cite{Agashe:2014kda}. The radius is defined from the derivative of the form factor at zero momentum square transferred point as
\begin{equation}
\langle r^2\rangle=-\frac{6}{F(0)}\frac{dF(Q^2)}{dQ^2}\bigg|_{Q^2=0}.
\end{equation}
The model results of the charge and the magnetic radii of the nucleons are listed in Table~\ref{radius} compared with the experimental values in Ref.~\cite{Agashe:2014kda}.

We need to emphasize that we calculate the flavor singlet axial charge which is interpreted as the contribution of quark helicities to the proton spin in the partonic language. Known as the ``proton spin crisis,'' it is often difficult to understand the helicity contribution at the high energy scale with other low energy baryon properties. In this study, by taking into account the quark mass effect, we obtain the value of the axial charge as 0.308, which is consistent with the recent experimental analysis $0.330\pm0.011(\textrm{theo.})\pm0.025(\textrm{exp.})\pm0.028(\textrm{evol.})$~\cite{Airapetian:2006vy}. If the quark mass is neglected, the value will reduce to zero, which is an ultrarelativistic limit. Therefore, the nonzero quark mass is important to describe hadronic spin structures.

\begin{table}
\caption{The magnetic moments of octet baryons. The nuclear magneton is defined as $\mu_N=e\hbar/2m_p$.\label{moments}}
\begin{tabular*}{0.5\textwidth}{m{0.12\textwidth}m{0.12\textwidth}c}
\hline\hline
 Baryon  & $\mu_{\textrm{th}}/\mu_N$ & $\mu_{\textrm{ex}}/\mu_N$\cite{Agashe:2014kda} \\
\hline
$p$ & \ \ $2.785$ & $2.792847356 \pm 0.000000023$ \\
$n$ & $-1.790$ & $-1.9130427 \pm 0.0000005$ \\
$\Lambda$ & $-0.799$ & $-0.613 \pm 0.004$ \\
$\Sigma^+$ & \ \ $2.414$ & $2.458 \pm 0.010$ \\
$\Sigma^-$ & $-0.896$ & $-1.160 \pm 0.025$ \\
$\Sigma^0\to\Lambda$ & \ \ $1.594$ & $1.61 \pm 0.08$ \\
$\Xi^0$ & $-1.181$ & $-1.250 \pm 0.014$ \\
$\Xi^-$ & $-0.589$ & $-0.6507 \pm 0.0025$ \\
\hline\hline
\end{tabular*}
\end{table}

\begin{table}
\caption{The charge and the magnetic radii of the proton and the neutron.\label{radius}}
\begin{tabular*}{0.5\textwidth}{m{0.12\textwidth}m{0.16\textwidth}c}
\hline\hline
& Theory value & Experiment value~\cite{Agashe:2014kda}\\
\hline
$r^p_E$ & $0.8819$\,fm & $0.8775 \pm 0.0051$\,fm\\
$r^p_M$ & $0.8783$\,fm & $0.777 \pm 0.016$\,fm\\
$\langle (r^n_E)^2\rangle$ & $-0.1390$\,fm$^2$ & $-0.1161 \pm 0.0022$\,fm$^2$\\
$r^n_M$ & $0.8392$\,fm & $0.862^{+0.009}_{-0.008}$\,fm\\
\hline\hline
\end{tabular*}
\end{table}

In summary, we investigate the octet and decuplet baryon properties in a light-front holographic soft-wall model. On the basis of the correspondence established on the limit of massless quarks, we include the corrections from nonzero masses of the constituents and obtain the modified LFWFs of both $L=0$ and $L=1$ states for baryons. Our results of baryon spectra, form factors, magnetic moments, and the charge and the magnetic radii match the experimental data well. The axial charge, which describes the quark helicity contribution to the nucleon spin, is also consistent with the analysis from the experimental data. Therefore, the AdS/CFT correspondence, or more generally the gauge/gravity duality, provides a powerful tool to study hadronic physics at both low and high energy scales, and if the effect from nonvanishing quark mass is taken into account, it is also successful in the study of spin-related issues.

\acknowledgments{This work is supported by the National Natural Science Foundation of China (Grants No.~11035003, No.~11120101004 and No.~11475006).}

\section*{}
{\it Note added.---} Some phenomenological baryon LFWFs based on the soft-wall model with finite quark mass were also proposed in Ref.~\cite{Gutsche:2013zia} and were applied to investigate nucleon structures.


\begin{thebibliography}{}


%\cite{Brodsky:2014yha}
\bibitem{Brodsky:2014yha}
  S.~J.~Brodsky, G.~F.~de Teramond, H.~G.~Dosch, and J.~Erlich,
  %``Light-Front Holographic QCD and Emerging Confinement,''
  Phys.\ Rep.\  {\bf 584}, 1 (2015).
  %%[arXiv:1407.8131 [hep-ph]].
  %%CITATION = ARXIV:1407.8131;%%


%\cite{Maldacena:1997re}
\bibitem{Maldacena:1997re}
  J.~M.~Maldacena,
  %``The Large N limit of superconformal field theories and supergravity,''
  Int.\ J.\ Theor.\ Phys.\  {\bf 38}, 1113 (1999)
  [Adv.\ Theor.\ Math.\ Phys.\  {\bf 2}, 231 (1998)].
  %%CITATION = HEP-TH/9711200;%%
  %10729 citations counted in INSPIRE as of 24 Apr 2015


\bibitem{Gubser:1998bc}
  S.~S.~Gubser, I.~R.~Klebanov, and A.~M.~Polyakov,
  %``Gauge theory correlators from non-critical string theory,''
  Phys.\ Lett.\  B {\bf 428}, 105 (1998).
  %%CITATION = PHLTA,B428,105;%%


\bibitem{Witten:1998qj}
  E.~Witten,
  %``Anti-de Sitter space and holography,''
  Adv.\ Theor.\ Math.\ Phys.\  {\bf 2}, 253 (1998).
  %%CITATION = 00203,2,253;%%


%\cite{Gross:1973ju}
\bibitem{Gross:1973ju}
  D.~J.~Gross and F.~Wilczek,
  %``Asymptotically Free Gauge Theories. 1,''
  Phys.\ Rev.\ D {\bf 8}, 3633 (1973);
  %%CITATION = PHRVA,D8,3633;%%
  %1970 citations counted in INSPIRE as of 24 Apr 2015
%\cite{Gross:1973id}
%\bibitem{Gross:1973id}
  D.~J.~Gross and F.~Wilczek,
  %``Ultraviolet Behavior of Nonabelian Gauge Theories,''
  Phys.\ Rev.\ Lett.\  {\bf 30}, 1343 (1973);
  %%CITATION = PRLTA,30,1343;%%
  %3599 citations counted in INSPIRE as of 24 Apr 2015
%\cite{Politzer:1973fx}
%\bibitem{Politzer:1973fx}
  H.~D.~Politzer,
  %``Reliable Perturbative Results for Strong Interactions?,''
  Phys.\ Rev.\ Lett.\  {\bf 30}, 1346 (1973).
  %%CITATION = PRLTA,30,1346;%%
  %3549 citations counted in INSPIRE as of 24 Apr 2015


%\cite{Furui:2006py}
\bibitem{Furui:2006py}
  S.~Furui and H.~Nakajima,
  %``Infrared features of unquenched finite temperature lattice Landau gauge QCD,''
  Phys.\ Rev.\ D {\bf 76}, 054509 (2007).
  %%CITATION = HEP-LAT/0612009;%%
  %26 citations counted in INSPIRE as of 24 Apr 2015


\bibitem{von Smekal:1997is}
  L.~von Smekal,  A.~Hauck, and R.~Alkofer,
  %``The Infrared behavior of gluon and ghost propagators in Landau gauge QCD,''
  Phys.\ Rev.\ Lett.\  {\bf 79}, 3591 (1997);
  %%CITATION = HEP-PH/9705242;%%
  %369 citations counted in INSPIRE as of 24 Apr 2015
%\cite{Binosi:2009qm}
%\bibitem{Binosi:2009qm}
  D.~Binosi and J.~Papavassiliou,
  %``Pinch Technique: Theory and Applications,''
  Phys.\ Rep.\  {\bf 479}, 1 (2009).
  %[arXiv:0909.2536 [hep-ph]].
  %%CITATION = ARXIV:0909.2536;%%
  %209 citations counted in INSPIRE as of 24 Apr 2015


%\cite{Deur:2005cf}
\bibitem{Deur:2005cf}
  A.~Deur, V.~Burkert, J.-P.~Chen, and W.~Korsch,
  %``Experimental determination of the effective strong coupling constant,''
  Phys.\ Lett.\ B {\bf 650}, 244 (2007);
  %[hep-ph/0509113].
  %%CITATION = HEP-PH/0509113;%%
  %72 citations counted in INSPIRE as of 24 Apr 2015
%\cite{Deur:2008rf}
%\bibitem{Deur:2008rf}
  A.~Deur, V.~Burkert, J.-P.~Chen, and W.~Korsch,
  %``Determination of the effective strong coupling constant alpha(s,g(1))(Q**2) from CLAS spin structure function data,''
  Phys.\ Lett.\ B {\bf 665}, 349 (2008).
  %[arXiv:0803.4119 [hep-ph]].
  %%CITATION = ARXIV:0803.4119;%%
  %69 citations counted in INSPIRE as of 24 Apr 2015


%\cite{Brodsky:2008be}
\bibitem{Brodsky:2008be}
  S.~J.~Brodsky and R.~Shrock,
  %``Maximum Wavelength of Confined Quarks and Gluons and Properties of Quantum Chromodynamics,''
  Phys.\ Lett.\ B {\bf 666}, 95 (2008).
  %[arXiv:0806.1535 [hep-th]].
  %%CITATION = ARXIV:0806.1535;%%
  %95 citations counted in INSPIRE as of 24 Apr 2015


%\cite{Karch:2006pv}
\bibitem{Karch:2006pv}
  A.~Karch, E.~Katz, D.~T.~Son, and M.~A.~Stephanov,
  %``Linear confinement and AdS/QCD,''
  Phys.\ Rev.\ D {\bf 74}, 015005 (2006).
  %[hep-ph/0602229].
  %%CITATION = HEP-PH/0602229;%%
  %568 citations counted in INSPIRE as of 24 Apr 2015


%\cite{Polchinski:2001tt}
\bibitem{Polchinski:2001tt}
  J.~Polchinski and M.~J.~Strassler,
  %``Hard scattering and gauge / string duality,''
  Phys.\ Rev.\ Lett.\  {\bf 88}, 031601 (2002).
  %[hep-th/0109174].
  %%CITATION = HEP-TH/0109174;%%
  %528 citations counted in INSPIRE as of 21 May 2015


%\cite{CasalderreySolana:2011us}
\bibitem{CasalderreySolana:2011us}
  J.~Casalderrey-Solana, H.~Liu, D.~Mateos, K.~Rajagopal, and U.~A.~Wiedemann,
  %``Gauge/String Duality, Hot QCD and Heavy Ion Collisions,''
  arXiv:1101.0618.
  %%CITATION = ARXIV:1101.0618;%%
  %303 citations counted in INSPIRE as of 21 May 2015


%\cite{Brodsky:2003px}
\bibitem{Brodsky:2003px}
  S.~J.~Brodsky and G.~F.~de Teramond,
  %``Light-front hadron dynamics and AdS/CFT correspondence,''
  Phys.\ Lett.\ B {\bf 582}, 211 (2004).
  %[hep-th/0310227].
  %%CITATION = HEP-TH/0310227;%%
  %193 citations counted in INSPIRE as of 24 Apr 2015


%\cite{Brodsky:2013ar}
\bibitem{Brodsky:2013ar}
  S.~J.~Brodsky, G.~F.~De Teramond, and H.~G.~Dosch,
  %``Threefold Complementary Approach to Holographic QCD,''
  Phys.\ Lett.\ B {\bf 729}, 3 (2014).
  %[arXiv:1302.4105 [hep-th]].
  %%CITATION = ARXIV:1302.4105;%%
  %36 citations counted in INSPIRE as of 24 Apr 2015


%\cite{Brodsky:1997de}
\bibitem{Brodsky:1997de}
  S.~J.~Brodsky, H.~C.~Pauli, and S.~S.~Pinsky,
  %``Quantum chromodynamics and other field theories on the light cone,''
  Phys.\ Rep.\  {\bf 301}, 299 (1998).
  %[hep-ph/9705477].
  %%CITATION = HEP-PH/9705477;%%
  %918 citations counted in INSPIRE as of 24 Apr 2015


%\cite{Dirac:1949cp}
\bibitem{Dirac:1949cp}
  P.~A.~M.~Dirac,
  %``Forms of Relativistic Dynamics,''
  Rev.\ Mod.\ Phys.\  {\bf 21}, 392 (1949).
  %%CITATION = RMPHA,21,392;%%
  %1415 citations counted in INSPIRE as of 24 Apr 2015


%\cite{Brodsky:2007hb}
\bibitem{Brodsky:2007hb}
  S.~J.~Brodsky and G.~F.~de Teramond,
  %``Light-Front Dynamics and AdS/QCD Correspondence: The Pion Form Factor in the Space- and Time-Like Regions,''
  Phys.\ Rev.\ D {\bf 77}, 056007 (2008).
  %[arXiv:0707.3859 [hep-ph]].
  %%CITATION = ARXIV:0707.3859;%%
  %261 citations counted in INSPIRE as of 24 Apr 2015


%\cite{Brodsky:2008pf}
\bibitem{Brodsky:2008pf}
  S.~J.~Brodsky and G.~F.~de Teramond,
  %``Light-Front Dynamics and AdS/QCD Correspondence: Gravitational Form Factors of Composite Hadrons,''
  Phys.\ Rev.\ D {\bf 78}, 025032 (2008).
  %[arXiv:0804.0452 [hep-ph]].
  %%CITATION = ARXIV:0804.0452;%%
  %91 citations counted in INSPIRE as of 24 Apr 2015


%\cite{Brodsky:2006uqa}
\bibitem{Brodsky:2006uqa}
  S.~J.~Brodsky and G.~F.~de Teramond,
  %``Hadronic spectra and light-front wavefunctions in holographic QCD,''
  Phys.\ Rev.\ Lett.\  {\bf 96}, 201601 (2006).
  %[hep-ph/0602252].
  %%CITATION = HEP-PH/0602252;%%
  %266 citations counted in INSPIRE as of 24 Apr 2015


%\cite{deTeramond:2008ht}
\bibitem{deTeramond:2008ht}
  G.~F.~de Teramond and S.~J.~Brodsky,
  %``Light-Front Holography: A First Approximation to QCD,''
  Phys.\ Rev.\ Lett.\  {\bf 102}, 081601 (2009).
  %[arXiv:0809.4899 [hep-ph]].
  %%CITATION = ARXIV:0809.4899;%%
  %172 citations counted in INSPIRE as of 24 Apr 2015


%\cite{Kirsch:2006he}
\bibitem{Kirsch:2006he}
  I.~Kirsch,
  %``Spectroscopy of fermionic operators in AdS/CFT,''
  J.\ High Energy Phys. {\bf 09} (2006) 052.
  %[hep-th/0607205].
  %%CITATION = HEP-TH/0607205;%%
  %39 citations counted in INSPIRE as of 24 Apr 2015


%\cite{deTeramond:2013it}
\bibitem{deTeramond:2013it}
  G.~F.~de Teramond, H.~G.~Dosch, and S.~J.~Brodsky,
  %``Kinematical and Dynamical Aspects of Higher-Spin Bound-State Equations in Holographic QCD,''
  Phys.\ Rev.\ D {\bf 87}, 075005 (2013).
  %[arXiv:1301.1651 [hep-ph]].
  %%CITATION = ARXIV:1301.1651;%%
  %29 citations counted in INSPIRE as of 24 Apr 2015


%\cite{Brodsky:2008pg}
\bibitem{Brodsky:2008pg}
  S.~J.~Brodsky and G.~F.~de Teramond,
  %``AdS/CFT and Light-Front QCD,''
  Proceedings of the International School of Subnuclear Physics, Vol.\ {\bf 45}, 139 (2009),
  arXiv:0802.0514.
  %%CITATION = ARXIV:0802.0514;%%
  %120 citations counted in INSPIRE as of 24 Apr 2015


%\cite{Abidin:2009hr}
\bibitem{Abidin:2009hr}
  Z.~Abidin and C.~E.~Carlson,
  %``Nucleon electromagnetic and gravitational form factors from holography,''
  Phys.\ Rev.\ D {\bf 79}, 115003 (2009).
  %[arXiv:0903.4818 [hep-ph]].
  %%CITATION = ARXIV:0903.4818;%%
  %65 citations counted in INSPIRE as of 24 Apr 2015


%\cite{deTeramond:2014asa}
\bibitem{deTeramond:2014asa}
  G.~F.~de Teramond, H.~G.~Dosch, and S.~J.~Brodsky,
  %``Baryon Spectrum from Superconformal Quantum Mechanics and its Light-Front Holographic Embedding,''
  Phys.\ Rev.\ D {\bf 91}, 045040 (2015);
  %%[arXiv:1411.5243 [hep-ph]].
  %%CITATION = ARXIV:1411.5243;%%
%\cite{Dosch:2015nwa}
%\bibitem{Dosch:2015nwa}
  H.~G.~Dosch, G.~F.~de Teramond, and S.~J.~Brodsky,
  %``Superconformal Baryon-Meson Symmetry and Light-Front Holographic QCD,''
  Phys.\ Rev.\ D {\bf 91}, 085016 (2015).
  %%[arXiv:1501.00959 [hep-th]].
  %%CITATION = ARXIV:1501.00959;%%


%\cite{Wigner:1939cj}
\bibitem{Wigner:1939cj}
  E.~P.~Wigner,
  %``On Unitary Representations of the Inhomogeneous Lorentz Group,''
  Annals Math.\  {\bf 40}, 149 (1939)
  [Nucl.\ Phys.\ Proc.\ Suppl.\  {\bf 6}, 9 (1989)].
  %%CITATION = ANMAA,40,149;%%
  %1054 citations counted in INSPIRE as of 24 Apr 2015


%\cite{Ma:1991xq}
\bibitem{Ma:1991xq}
  B.-Q.~Ma,
  %``Melosh rotation: Source for the proton's missing spin,''
  J.\ Phys.\ G {\bf 17}, L53 (1991);
  %[arXiv:0711.2335 [hep-ph]].
  %%CITATION = ARXIV:0711.2335;%%
  %81 citations counted in INSPIRE as of 24 Apr 2015
%\cite{Ma:1992sj}
%\bibitem{Ma:1992sj}
  B.-Q.~Ma and Q.-R.~Zhang,
  %``The Proton spin and the Wigner rotation,''
  Z.\ Phys.\ C {\bf 58}, 479 (1993).
  %[hep-ph/9306241].
  %%CITATION = HEP-PH/9306241;%%
  %89 citations counted in INSPIRE as of 24 Apr 2015


%\cite{Brodsky:1980zm}
\bibitem{Brodsky:1980zm}
  S.~J.~Brodsky and S.~D.~Drell,
  %``The Anomalous Magnetic Moment and Limits on Fermion Substructure,''
  Phys.\ Rev.\ D {\bf 22}, 2236 (1980).
  %%CITATION = PHRVA,D22,2236;%%
  %456 citations counted in INSPIRE as of 24 Apr 2015


%\cite{Grigoryan:2007my}
\bibitem{Grigoryan:2007my}
  H.~R.~Grigoryan and A.~V.~Radyushkin,
  %``Structure of vector mesons in holographic model with linear confinement,''
  Phys.\ Rev.\ D {\bf 76}, 095007 (2007).
  %[arXiv:0706.1543 [hep-ph]].
  %%CITATION = ARXIV:0706.1543;%%
  %114 citations counted in INSPIRE as of 24 Apr 2015


%\cite{Sachs:1962zzc}
\bibitem{Sachs:1962zzc}
  R.~G.~Sachs,
  %``High-Energy Behavior of Nucleon Electromagnetic Form Factors,''
  Phys.\ Rev.\  {\bf 126}, 2256 (1962).
  %%CITATION = PHRVA,126,2256;%%
  %181 citations counted in INSPIRE as of 24 Apr 2015


%\cite{Arrington:2007ux}
\bibitem{Arrington:2007ux}
  J.~Arrington, W.~Melnitchouk, and J.~A.~Tjon,
  %``Global analysis of proton elastic form factor data with two-photon exchange corrections,''
  Phys.\ Rev.\ C {\bf 76}, 035205 (2007).
  %[arXiv:0707.1861 [nucl-ex]].
  %%CITATION = ARXIV:0707.1861;%%
  %193 citations counted in INSPIRE as of 24 Apr 2015


%\cite{Walker:1993vj}
\bibitem{Walker:1993vj}
  R.~C.~Walker {\it et al.},
  %``Measurements of the proton elastic form-factors for 1-GeV/c**2 <= Q**2 <= 3-GeV/C**2 at SLAC,''
  Phys.\ Rev.\ D {\bf 49}, 5671 (1994).
  %%CITATION = PHRVA,D49,5671;%%
  %249 citations counted in INSPIRE as of 24 Apr 2015


%\cite{Borkowski:1974mb}
\bibitem{Borkowski:1974mb}
  F.~Borkowski, G.~G.~Simon, V.~H.~Walther, and R.~D.~Wendling,
  %``Electromagnetic Form-Factors of the Proton at Low Four-Momentum Transfer,''
  Nucl.\ Phys.\ {\bf B93}, 461 (1975).
  %%CITATION = NUPHA,B93,461;%%
  %129 citations counted in INSPIRE as of 24 Apr 2015


%\cite{Kubon:2001rj}
\bibitem{Kubon:2001rj}
  G.~Kubon {\it et al.},
  %``Precise neutron magnetic form-factors,''
  Phys.\ Lett.\ B {\bf 524}, 26 (2002).
  %[nucl-ex/0107016].
  %%CITATION = NUCL-EX/0107016;%%
  %155 citations counted in INSPIRE as of 24 Apr 2015


%\cite{Glazier:2004ny}
\bibitem{Glazier:2004ny}
  D.~I.~Glazier {\it et al.},
  %``Measurement of the electric form-factor of the neutron at Q**2 = 0.3-(GeV/c)**2 to 0.8-(GeV/c)**2,''
  Eur.\ Phys.\ J.\ A {\bf 24}, 101 (2005).
  %[nucl-ex/0410026].
  %%CITATION = NUCL-EX/0410026;%%
  %98 citations counted in INSPIRE as of 24 Apr 2015


%\cite{Markowitz:1993hx}
\bibitem{Markowitz:1993hx}
  P.~Markowitz {\it et al.},
  %``Measurement of the magnetic form-factor of the neutron,''
  Phys.\ Rev.\ C {\bf 48}, R5 (1993).
  %%CITATION = PHRVA,C48,5;%%
  %127 citations counted in INSPIRE as of 24 Apr 2015


%\cite{Bruins:1995ns}
\bibitem{Bruins:1995ns}
  E.~E.~W.~Bruins {\it et al.},
  %``Measurement of the neutron magnetic form-factor,''
  Phys.\ Rev.\ Lett.\  {\bf 75}, 21 (1995).
  %%CITATION = PRLTA,75,21;%%
  %153 citations counted in INSPIRE as of 24 Apr 2015


%\cite{Lung:1992bu}
\bibitem{Lung:1992bu}
  A.~Lung {\it et al.},
  %``Measurements of the electric and magnetic form-factors of the neutron from Q**2 = 1.75-GeV/c**2 to 4-GeV/c**2,''
  Phys.\ Rev.\ Lett.\  {\bf 70}, 718 (1993).
  %%CITATION = PRLTA,70,718;%%
  %227 citations counted in INSPIRE as of 24 Apr 2015


%\cite{Zhu:2001md}
\bibitem{Zhu:2001md}
  H.~Zhu {\it et al.}  (E93026 Collaboration),
  %``A Measurement of the electric form-factor of the neutron through polarized-d (polarized-e, e-prime n)p at Q**2 = 0.5-(GeV/c)**2,''
  Phys.\ Rev.\ Lett.\  {\bf 87}, 081801 (2001).
  %[nucl-ex/0105001].
  %%CITATION = NUCL-EX/0105001;%%
  %160 citations counted in INSPIRE as of 24 Apr 2015


%\cite{Warren:2003ma}
\bibitem{Warren:2003ma}
  G.~Warren {\it et al.}  (Jefferson Lab E93-026 Collaboration),
  %``Measurement of the electric form-factor of the neutron at $Q^2$ = 0.5 and 1.0 $GeV^2/c^2$,''
  Phys.\ Rev.\ Lett.\  {\bf 92}, 042301 (2004).
  %[nucl-ex/0308021].
  %%CITATION = NUCL-EX/0308021;%%
  %115 citations counted in INSPIRE as of 24 Apr 2015


%\cite{Rohe:1999sh}
\bibitem{Rohe:1999sh}
  D.~Rohe {\it et al.},
  %``Measurement of the neutron electric form-factor G(en) at 0.67-(GeV/c)**2 via He-3(pol.)(e(pol.),e' n),''
  Phys.\ Rev.\ Lett.\  {\bf 83}, 4257 (1999).
  %%CITATION = PRLTA,83,4257;%%
  %163 citations counted in INSPIRE as of 24 Apr 2015


%\cite{Agashe:2014kda}
\bibitem{Agashe:2014kda}
  K.~A.~Olive {\it et al.}  (Particle Data Group),
  %``Review of Particle Physics,''
  Chin.\ Phys.\ C {\bf 38}, 090001 (2014).
  %%CITATION = CHPHD,C38,090001;%%
  %1018 citations counted in INSPIRE as of 24 Apr 2015


%\cite{Airapetian:2006vy}
\bibitem{Airapetian:2006vy}
  A.~Airapetian {\it et al.}  (HERMES Collaboration),
  %``Precise determination of the spin structure function g(1) of the proton, deuteron and neutron,''
  Phys.\ Rev.\ D {\bf 75}, 012007 (2007).
  %[hep-ex/0609039].
  %%CITATION = HEP-EX/0609039;%%
  %307 citations counted in INSPIRE as of 24 Apr 2015

%\cite{Gutsche:2013zia}
\bibitem{Gutsche:2013zia}
  T.~Gutsche, V.~E.~Lyubovitskij, I.~Schmidt, and A.~Vega,
  %``Light-front quark model consistent with Drell-Yan-West duality and quark counting rules,''
  Phys.\ Rev.\ D {\bf 89}, 054033 (2014);
%%  [Phys.\ Rev.\ D {\bf 92}, 019902 (2015)]
%%  [arXiv:1306.0366 [hep-ph]].
  %%CITATION = ARXIV:1306.0366;%%
%\cite{Gutsche:2014yea}
%\bibitem{Gutsche:2014yea}
  T.~Gutsche, V.~E.~Lyubovitskij, I.~Schmidt, and A.~Vega,
  %``Nucleon structure in a light-front quark model consistent with quark counting rules and data,''
  Phys.\ Rev.\ D {\bf 91}, 054028 (2015).
%%  [arXiv:1411.1710 [hep-ph]].
  %%CITATION = ARXIV:1411.1710;%%

\end{thebibliography}
\end{document}